\input harvmac.tex
\input epsf
\noblackbox

\def\zp{z^\prime}
\def\cC{{\cal C}}
\def\IP{{\bf P}}
\def\ff{\bf F}
\def\xthree{X^{(3)}}
\def\xfour{X^{(4)}}
\def\xfive{X^{(5)}}
\def\bb{{\bf B}}

\def\bfone{\relax{\rm 1\kern-.35em 1}}
\def\inbar{\vrule height1.5ex width.4pt depth0pt}
\def\IC{\relax\,\hbox{$\inbar\kern-.3em{\mss C}$}}
\def\ID{\relax{\rm I\kern-.18em D}}
\def\IF{\relax{\rm I\kern-.18em F}}
\def\IH{\relax{\rm I\kern-.18em H}}
\def\II{\relax{\rm I\kern-.17em I}}
\def\IN{\relax{\rm I\kern-.18em N}}
\def\IQ{\relax\,\hbox{$\inbar\kern-.3em{\rm Q}$}}
\def\us#1{\underline{#1}}
\def\IR{\relax{\rm I\kern-.18em R}}
\font\cmss=cmss10 \font\cmsss=cmss10 at 7pt
\def\ZZ{\relax\ifmmode\mathchoice
{\hbox{\cmss Z\kern-.4em Z}}{\hbox{\cmss Z\kern-.4em Z}}
{\lower.9pt\hbox{\cmsss Z\kern-.4em Z}}
{\lower1.2pt\hbox{\cmsss Z\kern-.4em Z}}\else{\cmss Z\kern-.4em
Z}\fi}

\def\cC{{\cal C}}

\def\cL{{\cal L}}

\def\nup#1({Nucl.\ Phys.\ $\us {B#1}$\ (}
\def\plt#1({Phys.\ Lett.\ $\us  {B#1}$\ (}
\def\cmp#1({Comm.\ Math.\ Phys.\ $\us  {#1}$\ (}
\def\prp#1({Phys.\ Rep.\ $\us  {#1}$\ (}
\def\prl#1({Phys.\ Rev.\ Lett.\ $\us  {#1}$\ (}
\def\prv#1({Phys.\ Rev.\ $\us  {#1}$\ (}
\def\mpl#1({Mod.\ Phys.\ Let.\ $\us  {A#1}$\ (}
\def\ijmp#1({Int.\ J.\ Mod.\ Phys.\ $\us{A#1}$\ (}
\def\tit#1|{{\it #1},\ }

\def\Coe#1.#2.{{#1\over #2}}

\def\coe#1.#2.{\relax{\textstyle {#1 \over #2}}\displaystyle}

\def\br{\hfill\break}

\overfullrule=0pt

%
%
%
\Title{\vbox{
\hbox{CERN-TH/96-184}
\hbox{HUTP-96/A031}
\hbox{\tt hep-th/9607139}
}}{BPS States of Exceptional Non-Critical Strings}

\bigskip
\centerline{Albrecht Klemm$^{a}$, Peter Mayr$^{a}$ and Cumrun Vafa$^{b}$}
\bigskip
\bigskip\centerline{\it $^{a}$Theory Division, CERN, 1211 Geneva 23,
Switzerland}
\centerline{\it $^{b}$Lyman Laboratory of Physics, Harvard
University,
Cambridge, MA 02138}

\vskip .3in
We study the BPS states of non-critical strings which arise for
zero size instantons of exceptional groups.   This is accomplished by using
F-theory and M-theory duals and by employing mirror symmetry to compute the
degeneracy of membranes wrapped
around 2-cycles of the Calabi--Yau threefold.  We find evidence for 
a number of novel
physical phenomena, including having infinitely many light states
with the first lightest state including a nearly massless gravitino.

\vskip 3cm
\centerline{\it To appear in the proceedings of the  conference}
\centerline{\it  "Advanced Quantum
Field Theory''  (in memory of Claude Itzykson)}
\goodbreak

\Date{\vbox{\hbox{\sl {July 1996}}}}

%
\parskip=4pt plus 15pt minus 1pt
\baselineskip=15pt plus 2pt minus 1pt

\newsec{Introduction}
One of the most important aspects in the recent developments
in string dualities has been the fact that string compactifications
which classically look singular have often a reinterpretation
in terms of light solitonic states.  These light solitonic states
can in general have $p$-spatial coordinates whose
`tension' $T_p$ goes to zero.  In general if there are many such
objects the most relevant light states correspond to the one for which
the relevant mass parameter
$T_p^{1\over p+1}$ is the smallest \ref\witcom{E. Witten, {\sl Some
Comments on String Dynamics}, hep-th/9507121}\ref\hul{C. Hull, {\sl String
Dynamics at Strong Coupling}, \nup468(1996)113}.  
Typically
in string theory when there are multitude of light states all 
$T_p\propto \epsilon$ as $\epsilon \rightarrow 0$, and thus the lightest states
correspond to the solitons with the smallest value of $p$.
For example for type IIA and type IIB strings the relevant
light states typically correspond to D-branes wrapped around vanishing cycles
whose volume is proportional to $\epsilon$.
Given the fact that the type IIA (type IIB) has
even (odd) D-branes \ref\pol{J. Polchinski,
{\sl  Dirichlet  Branes and Ramond-Ramond Charges},
\prl75(1995)4724, hep-th/9510017}\ we see that the most relevant
lightest states are either massless particles ($p=0$) or
tensionless strings ($p=1$).

The case of massless solitonic particles is easier to understand
from the viewpoint of quantum field theories.  Examples
of this class include type IIA compactifications near an
ADE singularity of $K3$ where the resulting solitons are responsible
for the enhanced ADE gauge symmetry \ref\wisd{E. Witten, {\sl String Theory
Dynamics in Various Dimensions}, \nup443(1995)85}\
or type IIB near a conifold
singularity of a Calabi-Yau threefold which leads to a
massless hypermultiplet
\ref\stbh{A. Strominger, {\sl Massless Black Holes
and Conifolds in String Theory}, \nup451(1995)95}.
Another example of this is when $SO(32)$ instantons shrink to zero
size which leads to enhanced gauge symmetry with some matter multiplets
\ref\witin{E. Witten, {\sl Small Instantons in String Theory},
\nup460(1996)541}.  This is not to say that all the cases with
massless solitonic particles are easy to understand
in terms of local Lagrangians, as for example
one can encounter simultaneous massless electric and magnetic particles
\ref\ard{P.C. Argyres and M. Douglas, {\sl
New Phenomena in $SU(3)$ Sypersymmetric Gauge Theory},
\nup448(1995)93}.

The case with $p=1$ is more unfamiliar, and probably
physically more interesting, as it may signal the
appearance of new critical quantum field theories based on strings
rather than particles, even if one ignores gravitational effects.
An example of this includes type IIB near an ADE singularity of $K3$
\witcom\ref\witst{N. Seiberg and E. Witten, {\sl
Comments on String Dynamics in Six-Dimensions},
\nup471 (1996) 121, hep-th/9603003}.  Given our relative lack of familiarity
with quantum field theories based on loops it is natural
to try to get a first order understanding of these theories by
relating them to the cases where the relevant light
degrees of freedom are again particles.  An attempt in this
direction is to further compactify on a circle.  In the resulting
theory the most relevant lightest states are now again particles
corresponding to the tensionless string wrapped around the circle.
In this way for example the physics of type IIB near an ADE singularity times
a circle becomes understandable as an ordinary Higgs mechanism
in the resulting five dimensional theory.

{}From this analysis one may get the wrong impression that
by a further compactification on a circle we will always
obtain a situation with `simple physics' in one lower
dimension.  This turns out not to be the case.
Roughly speaking what happens sometimes
is that in the one lower dimensional
theory there are infinitely many massless particles interacting
with a tensionless string.   At first sight this may appear puzzling
as a string wrapped around a circle does not seem to have infinitely
many light degrees of freedom in store.  Moreover it appears
to be more relevant than the unwrapped string.
 Technically the way
this comes about is that due to a quantum effect the wrapped
string
has a smaller tension  than the unwrapped string and thus
as the wrapped string becomes tensionless, the unwrapped
string still has a positive tension.  However one can pass
through this transition
point beyond which the wrapped string formally acquires a {\it negative}
tension.  As we hit the second, and more interesting, transition point
the unwrapped string becomes tensionless.  Also at this point
infinitely many of the BPS states of the wrapped string
become massless.
  Moreover the tensionless string is as relevant
as these massless particles. Thus the situation
appears roughly as a tensionless string interacting with
infinitely many light particles resulting from wrapped
states of the same string which due to quantum
effects has a negative tension.
{}From the fact that after the first transition and before the
second transition the wrapped string acquires a negative tension,
one see that one needs a better ``dual picture'' and this turns out
to be provided by the geometry of special singularities of
 Calabi--Yau manifolds in the context of M-theory.
A study of an example of this situation is one of the main aims of
the present paper.

The class of theories we study correspond to $N=1$ theories in $d=6$,
as is the case for instance for $E_8\times E_8$
 heterotic strings compactified on $K3$.
It is natural to wonder what happens
when an $E_8$ instanton shrinks to zero size.
In this case
one can use M-theory description of $E_8\times E_8$ heterotic
string \ref\hw{P. Horava and E. Witten, {\sl Heterotic
and Type I  String Dynamics from Eleven-Dimensions},
\nup460(1996)541}\ to gain insight into
the nature of the singularity \ref\gh{O. Ganor and
A. Hanany, {\sl Small $E(8)$ Instantons and
Tensionless  Non-critical Strings}, \nup474 (1996) 122, hep-th/9602120}:
As an $E_8$ instanton shrinks it can be locally represented
as a 5-brane residing on the 9-brane boundary corresponding
to the $E_8$ whose instanton is shrinking in size.
If we deform this situation by moving the 5-brane off
the 9-brane one can see there is a tensionless string:  Since
the membrane of M-theory can end on both the 9-brane
\hw\ and the 5-brane
\ref\st{A. Strominger, {\sl Open P-Branes}, \plt 383 (1996) 44,
hep-th/9512059}\ref
\town{P. Townsend, {\sl D-Branes from M-Branes}, \plt373 (1996) 68},
in the limit the 5-brane and 9-brane touch we get a tensionless
string corresponding to the boundary of the membrane in
the common 6-dimensional space-time.

On the other hand heterotic strings on $K3$ are dual
to F-theory compactifications on Calabi-Yau threefolds
\ref\mvi{D. Morrison and C. Vafa, {\sl Compactifications of $F$ theory on
Calabi-Yau
Threefolds I}, \nup 473 (1996) 74, hep-th/9602114}\ref\mvii{D. Morrison
and C. Vafa,  {\sl Compactifications of $F$ theory on Calabi-Yau
Threefolds II}, \nup 476 (1996) 437, hep-th/9603161},
and it is possible to isolate
what corresponds to an $E_8$ instanton of zero size
\ref\wittra{E. Witten, {\sl Phase Transitions In $M$ Theory
And $F$ Theory}, \nup 471 (1996) 195, hep-th/9603150}\mvii .  We will use
this duality to study the spectrum of BPS states
for heterotic strings on $K3\times S^1$ when
an $E_8$ instanton shrinks to zero size and compare
it with the predictions of tensionless strings.
To do this study we need to count the number of solitonic
membranes of M-theory wrapped holomorphically around cycles
of the Calabi-Yau manifold.  We use mirror symmetry
to accomplish this.
We find perfect agreement with predictions
based on tensionless strings, subject to some very important
subtleties.   In particular the Calabi-Yau geometry refines
the description of the tensionless string suggesting certain quantum
corrections to the classical picture.
 We will also study the conjectured
duals for $E_d$ instantons of zero size \mvii\ and
find that they are naturally understandable as the $E_8$
non-critical string propagating in the presence
of Wilson lines breaking $E_8$ to $E_d$ (up to $U(1)$ factors).

The organization of this paper is as follows:
In section 2 we set up the predictions of BPS states
based on tensionless strings.  In section 3 we set up the
question of
the BPS spectrum on the type II side (in the context of
F-theory and M-theory ) in the context of counting curves in
del Pezzo surfaces sitting in the Calabi-Yau.
  In section 4 we compare
the predictions  and find agreement with expectations
based on the tensionless strings.  We also discuss some
aspects of the BPS spectrum on the type II
side which points towards new physics.  In an appendix we discuss some
technical aspects of the relevant singularity of the Calabi-Yau threefolds.

As we were completing this work an interesting paper appeared \ref\gan{J.
Ganor, {\sl
A Test of the Chiral $E(8)$ Current Algebra on a
6-D Non-critical String }, hep-th/9607020}\
which has some overlap with the present work.

\newsec{BPS states from non-critical $E_8$ string}
As mentioned in the last section in the context
of M-theory when a 5-brane meets
the 9-brane we have a situation dual to a small $E_8$ instanton
of heterotic string.  Actually, this is in the phase where
the instanton has shrunk to zero size and we have
`nucleated' a 5-brane which has departed from the 9-brane
world volume of the $E_8$.  A membrane stretched between
the 5-brane and 9-brane lives as a string on the common
6-dimensional space-time.  The tension for the string is proportional
to the distance between the 5-brane and the 9-brane.

To better understand the properties of this non-critical
string let us recall that if we have two parallel 9-branes
the resulting string will be the heterotic string, where
on each 9-brane lives an $E_8$ gauge symmetry, each inducing
a (say) left-moving $E_8$ current algebra on the string.  Recall also
that if we have two parallel 5-branes we get a
string in 6-dimension, which couples to an $N=2$ tensor
multiplet.  The degrees of freedom on this string is best
described by Green-Schwarz strings
 in 6 dimensions \ref\dvv{R. Dijkgraaf, E. and H. Verlinde,
{\sl BPS Quantization of the Five-Brane},
hep-th/9604055}\ref\sc{J. Schwarz, {\sl Self-Dual
Superstring in Six Dimensions}, hep-th/9604171}.
The light cone degrees of freedom are described by 4 left-moving
fields and 4 right-moving fields each transforming in the spinor
representation of the light cone group $O(4)$.

In the case at hand we have a non-critical string which has
half the supersymmetry of the above six dimensional string
resulting from stretched membrane
between the two 5-branes, as well as having half the $E_8$ current
algebra of the string resulting from a membrane
stretched between two 9-branes.  In fact
we have one left-moving $E_8$ current algebra at level one and one right-moving
spinor of $O(4)$ in the light cone gauge.  In addition we have,
in the light cone gauge the usual 4 transverse bosonic oscillators.

\subsec{Prediction for BPS States}
Now we further compactify on a circle and ask what are the BPS states
which carry a winding charge of this non-critical string?
This is a familiar situation encountered in the study of critical strings
\ref\dh{A. Dabholkar and J. Harvey, {\sl Nonrenormalisation of the Superstring
Tension},
\prl63(1989)478}.  In the case at hand we do not know enough
about the properties of the resulting non-critical string to rigorously
derive the spectrum of BPS states, but we will follow the same
line of argument as in the critical string case and derive
what seems to be the reasonable BPS spectrum of states:  Let
$(P_L,P_R)$ denote the left- and right-moving momenta of the string
on the resulting circle.  We have
$$(P_L,P_R)={1\over \sqrt{T}}({n\over 2R}-mR T,{n\over 2R}+mR T)$$
where $(n,m)$ denote the momentum and winding of the string
around the circle and $T$ denotes its tension.  The
overall factor of $1\over \sqrt{T}$ in front is put to make
$(P_L,P_R)$ dimensionless.
Note that
$${1\over 2}(P_R^2-P_L^2)=n\cdot m\ .$$
Let $(L_0,{\overline L}_0)$ denote the left- and right-moving
Hamiltonians corresponding to oscillating states
of the strings.  Since we are after BPS
states we restrict to ground state oscillator states for the
right-movers but arbitrary states on the left-movers.  We also need to
impose the equality of $L_0=\overline L_0$.  Assuming free oscillating
states we have
$$L_0={1\over 2 }P_L^2 +L_0^I+N$$
$${\overline L}_0={1\over 2}P_R^2+{\overline N}$$
where $L_0^I$ corresponds to the internal degrees of freedom
of the $E_8$ current algebra and $N$ denotes the contribution
of oscillators from 4 transverse bosonic states.  Similarly
${\overline N}$ denotes the oscillators contribution
from 4 bosonic and 4 fermionic right-moving oscillator states.
In principle there could have been a constant addition to
$L_0$ as is for instance for bosonic strings.  We will see
that in the case at hand to agree with predictions based on the
type II side we do not need any shifts.  To have a BPS state
we set $\overline N =0$ and we thus see that
\eqn\bpsc{N+{L_0^I}={1\over 2}(P_R^2-P_L^2)=n\cdot m}
The mass of the corresponding BPS state is given by
\eqn\mass{M=\big|\sqrt{T} P_R\big|=\big| {n\over 2R}+m RT \big|}
 With
more than one unit of winding, since we do not know enough
about the non-critical string, it is difficult to decide whether
we have stable new states at higher winding numbers, as would be
the case in critical string theories or that the multiply
wound state decays to singly wound states
(we shall find in section 4, using the type II dual that they indeed
do not form such bound states).
To avoid this complication we concentrate on the case $m=1$.
Setting $m=1$ in \mass\ we have
\eqn\mass{M_n=\big|\sqrt{T} P_R\big|=\big| {n\over 2R}+ RT \big|}
We see from \bpsc\ that if $d(n)$ denotes the degeneracy of
BPS states with $n$-units of momentum around the circle
with winding number one we have
$$q^{-{1\over 2}}\sum_{n=0}^{\infty} d(n)q^n={\chi_{E_8}(q)\over {\eta(q)}^4}
={\theta_{E_8}(q)\over {\eta(q)}^{12}}$$
where we have used the fact that the internal $L_0^I$ corresponds
to a level one $E_8$ Kac-Moody algebra and can be viewed in the bosonized
form as corresponding to 8 bosons compactified on the $E_8$ root lattice.
We thus have 12 oscillators 4 of which are space-time oscillators and
8 are internal and are scalar.  Note that we have thus learned that
$$d(0)=1, d(1)=252,d(2)=5130,...$$
with masses given by
$$M_0=RT,M_1=RT +{1\over R},M_2=RT +{2\over R},...$$
What about their space-time quantum numbers?  Again because
of our ignorance about non-critical strings it is difficult
to judge a priori what the right-moving
supersymmetric ground state should be.  Formally one would think
that the right-moving oscillator zero mode consists of a spinor of $O(4)$
in the light cone and so the space-time quantum
numbers of the right-moving ground state must be
 half a hypermultiplet which
transforms as $2(0,0)\oplus (1/2,0)$ of $O(4)$.  Together
with the inversely wound states, this would form a full hypermultiplet
as the ground state on the right-moving side.  To get the full quantum
number of the above BPS states we have to tensor
this with the left-moving quantum numbers.  There are two types
of quantum numbers, corresponding to space-time as well as $E_8$
representations.  The fact that they form $E_8$ representations
is natural when we recall that locally when an $E_8$ instanton
shrinks we expect at least {\it locally} to restore the $E_8$
and so all the states should form $E_8$ representations.
The $E_8$ content of the BPS states can be easily deduced
from the corresponding affine Kac-Moody degeneracies.
As for the space-time quantum numbers, we have four transverse
bosonic oscillators which transform as $({1\over 2},{1\over 2})$
of $O(4)$.  This thus allows us to find the quantum number of all
BPS states.  For example the state with $n=0$ is a hypermultiplet
singlet of $E_8$.  The state with $n=1$ is given by
\eqn\stw{\big[{\bf 248};4(0,0)\oplus ({1\over 2},0)\oplus (0,{1\over 2})\big]+
\big[{\bf 1};4({1\over 2},{1\over 2})\oplus (1,{1\over 2})\oplus
({1\over 2},1)\oplus (0, {1\over 2})\oplus ({1\over 2},0)\big]}
where the first entry denotes the $E_8$ content of the state.
Note that the spin of these  states goes all the way up to
spin $3/2$ (in four dimensional terms).  Similar decompositions
can be done for all higher values of $n$ as well.  Note that
as we start decreasing the tension towards zero the once wrapped
state with no momentum, which is the lightest state becomes a
massless hypermultiplet.  The rest of the BPS states, including
the one with $n=1$ are still massive in this limit.  So far we do not
see any exotic physics and, modulo the question of multiple
windings, the tensionless string seems to have
produced only a massless hypermultiplet.  As we will discuss
in the next section there is an important subtlety which is difficult
to see in this setup.  We will find that the interesting physics
is associated not with the point where the wrapped string becomes
tensionless with $T=0$ but actually when $T$
of the wrapped string becomes {\it negative}.
These are more clear from the viewpoints of F-theory and M-theory to which
we turn to in the next section.

It is natural to ask what happens when instantons for other
gauge groups shrink, for example that of other exceptional
groups $E_d$.   In the most standard form of the question
this issue does not naturally arise in the six dimensional
theories\foot{Actually in cases where we choose the instantons
in a sub-bundle of $E_8$ the other cases may also occur
in some special cases.},
because the only relevant gauge groups are $E_8\times E_8$
or $SO(32)$.  But the issue can naturally arise if we go to 5 dimensions.

A simple way to see this is to  suppose we start with the $E_8\times E_8$
heterotic strings in 10 dimensions and
 first go from 10 to 9 dimensions
on a circle where we turn on a Wilson line which breaks one
of the $E_8$'s to a smaller group\foot{
The following discussion is unaltered even
if we have other unbroken groups.  We use this
choice for later comparison with the F-theory conjecture
for such cases \mvii .}  say $E_d\times
U(1)^{8-d}$.
The choices for such Wilson lines are parametrized by $8-d$ real
parameters $A_i$ denoting the choice of the
Wilson line in each of the
$U(1)$'s. Now if we further compactify the heterotic string on $K3$
all the instantons will now reside in $E_d$. In such a case the
question of $E_d$ small instantons and their physical
interpretation arises.

We can use the picture of non-critical tensionless $E_8$ string
to gain insight into this situation as well.  The idea is
to consider the point at which the string is exactly tensionless
with $T=0$.  At this point the $E_8$ gauge symmetry is locally
restored as the $E_8$ instantons have zero size.  Now it
makes sense to turn on the Wilson lines and break $E_8
\rightarrow E_d \times U(1)^{8-d}$.  This situation seems to
be indistinguishable from having had started with
$E_d$ instantons and making their size shrink. Even though
this is not a proof it seems very plausible that the
two descriptions are identical.  If so then
we can learn about how the small $E_d$ instantons behave
as far as the spectrum of BPS states are concerned.  They
are simply the ones for $E_8$ strings deformed by the
Wilson lines on the circle.

Let us denote the relevant
Wilson lines as a vectors ${\bf W}_\alpha$, $\alpha=1\dots 8-d$ 
in the $E_8$
Cartan Lie algebra.  As noted above the BPS states
form representations of $E_8$.  Let ${\bf \Lambda}$ denote a
weight of a state of one of the BPS states.  Suppose
it originally had mass $M$.  After turning the Wilson line
the mass shifts in the usual way by
\eqn\wils{M\rightarrow M+\sum_\alpha {\bf \Lambda}\cdot {\bf W}_\alpha}
This now splits the BPS states into states which form
representations of $E_d\times U(1)^{8-d}$.

\newsec{F-theory and M-theory Descriptions}
We now analyze the same situation
from a dual viewpoint using the duality between heterotic string on $K3$
and F-theory
on elliptically fibered
Calabi-Yau threefolds \mvi \mvii .  The transition from a
5-brane approaching a 9-brane to having an instanton of finite
size is locally the same transition in the Calabi-Yau
language as going from an elliptically fibered Calabi--Yau
with base making a blow down from $\ff_1\rightarrow \IP^2$
\witin\mvii , where $\ff_1$ is the Hirzebruch surface
of degree 1.
 This turns out to be the same transition
which occurs for strong coupling for heterotic strings
with instanton numbers (11,13) (which is dual to F-theory
on Calabi--Yau over $\ff_1$) giving us another realization
of the transition we are considering.  At the transition
point an ``exceptional divisor'' $D$ which is a two sphere
with self-intersection -1 shrinks to zero size \mvi.  The tensionless
string in six dimensions is to be identified with the three brane
of type IIB wrapped around $D$ \witst.

There are two relevant K\"ahler classes for this transition
of the Calabi-Yau:  the K\"ahler class of $D$ which we denote
by $k_D$ and the K\"ahler class of the elliptic curve $k_E$.
Clearly the tension of the string in six dimensions $T\propto k_D$.
In the F-theory limit the K\"ahler class of the elliptic curve is
not dynamical and can be thought
of as formally being put to zero.
More precisely \ref\vef{C. Vafa, {\sl Evidence for F Theory}, \nup 496 (1996)
403, hep-th/9602065}\ if we compactify
further on a circle of radius $R$ down to five dimensions we have an
equivalence with M-theory on the same elliptic
Calabi-Yau with $k_E \propto 1/R$.  In the limit as $R\rightarrow \infty$
we obtain the F-theory compactification in six dimensions.

As it turns out the nature of the above transition is different between
6 and 5 dimensions (suggesting that there is a quantum correction
in the 5-dimensional theory) as has been elaborated in \mvii\ which
we will now review.
In the six dimensional case there is only one class $k_D$, as $k_E=0$ and the
transition takes place when $k_D=0$.  In the five dimensional theory
there are actually two transition points:  Let us fix the radius
(or equivalently fix $k_E$) and start decreasing $k_D$ to approach
the transition point (see Fig. 1).  When $k_D=0$,  $D$ has shrunk
to zero size and we have the tensionless string in 6 dimensions.
  But the interesting transition occurs further down
when we take $k_D<0$.  What this means is that we have done
a `flop' on $D$.  The actual transition point is when we reach
$k_D=-k_E$.  At this point an entire 4-cycle which is the
$E_8$ del Pezzo surface ($\IP^2$ blown up at 8 points)
which we denote by $\bb_8$ has shrunk
to zero size.
 In terms of the wrapped string this means
formally setting the tension to a negative value.   We identify
\eqn\ide{k_E={1\over 2R}\ , \ k_D=RT }
Note that there really are two physical transitions.  At the
first transition point where $k_D=0$ we have a 2-sphere shrunk
to zero size. In M-theory the membrane can wrap around
this two sphere and give rise to a massless
particle.  In fact this situation has already
been analyzed \wittra\
with the result that at this point one has a
single massless hypermultiplet.   In fact this is what
was expected based on the tensionless string description
discussed in the previous section where one obtains
a massless hypermultiplet.  However as we have already
emphasized the interesting transition is associated with a
different point at which at least formally
this string acquires a {\it negative tension}.  This is
also the point where the unwrapped string becomes
tensionless.
If we analytically
continue the formula obtained from the viewpoint of our original
string we see that among the BPS states we considered the string
which wraps around the circle {\it and} carries one unit
of momentum about the circle now becomes massless, i.e.
one would expect that at this transition point there are $252$
massless states.  Using the BPS counting of states
also for the type II side we will verify below
that this is indeed a correct supposition.
At this second
transition point we can wrap membranes about any 2-cycle on
$\bb_8$ and get a massless particle; the 252 states
just discussed should be among them.  At this point the unwrapped
tensionless string is associated with the 5-brane
of M-theory wrapped around $\bb_8$.
If we denote the volume of $\bb_8$ by $\epsilon^2$, then the
tension of the resulting string goes as $T\sim \epsilon^2$.
As far as dimensional
argument the resulting mass scale $T^{1/2}=\epsilon$ is the same
as the mass scale for the membranes wrapped around the 2-cycles
of $\bb_8$ and so are equally relevant \wittra .  If we compactify
further to 4 dimensions where we get equivalence with type IIA
theory on the same Calabi-Yau, we can have three different light
states: a 5-brane wrapped around vanishing 4-cycle,
a Dirichlet 4-brane wrapped around the vanishing 4-cycle
and finally a Dirichlet 2-brane wrapped around any vanishing
2-cycle in $\bb_8$.  It turns out that the simple dimensional
analysis now suggests that the relevant light states
are the 4-brane wrapped around vanishing 4-cycle which
lead to massless particles; thus there may
well be a local Lagrangian formulation of this theory
in the four dimensional case.  This should be very interesting
to identify.
\goodbreak\midinsert
\centerline{\epsfxsize 3.5truein\epsfbox{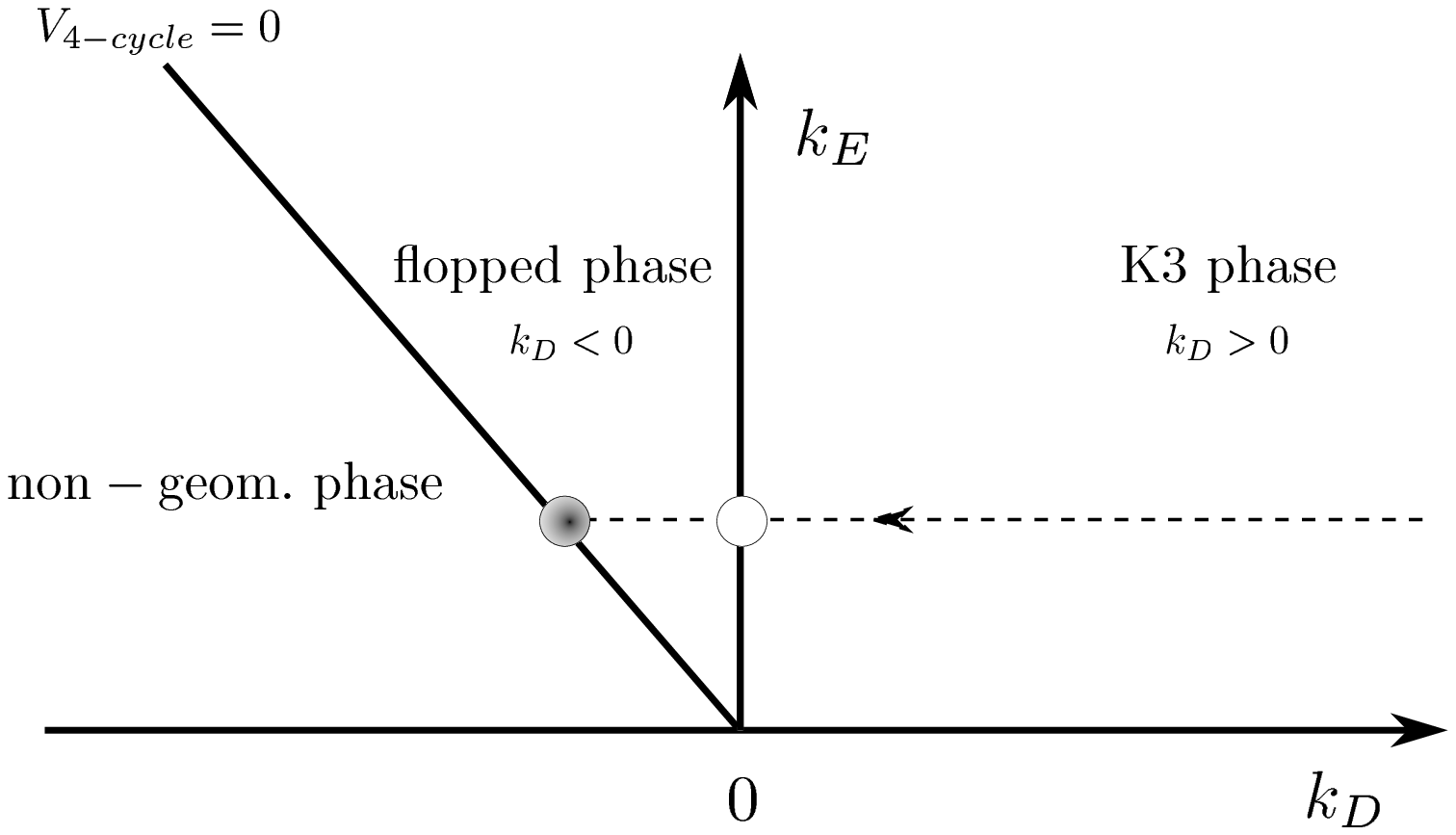}}\leftskip 2pc
\rightskip 2pc\noindent{\ninepoint\sl \baselineskip=8pt {\bf Fig.1}:
Phases of the K\"ahler moduli space. In 5d ($k_E \neq 0$) there is
a new phase where the string tension of the wrapped string
 $\sim k_D$ becomes formally
negative. Decreasing the tension of the wrapped
string following the dashed line
it becomes zero at the first transition point (white circle)
where one gets only a massless hypermultiplet from winding the string
around the circle. Continuing further to negative values one hits
the second transition point (black circle)
where the unwrapped string becomes tensionless.  In
addition there are infinitely many
massless point like objects arising from wrapped
states of the negative tension string interacting with the
 magnetically charged tensionless string.}\endinsert
Similarly it has been conjectured \mvii\ that the small $E_d$
instantons which one encounters in heterotic string compactifications
below 6 dimensions are dual to Calabi-Yau threefolds where a
del Pezzo surface of type $\bb_d$  (to be reviewed
in the next subsection) shrink to zero size.  In fact it should be
possible to derive this conjecture directly from the case of small
$E_8$ instantons following the physical idea of turning on
Wilson lines discussed at the end of the next section and
following the parallel geometric description of this operation
(as has been done in similar contexts in \ref\beret{
M. Bershadsky, K. Intriligator, S. Kachru,
D.R. Morrison, V. Sadov, C. Vafa, {\sl Geometric
Singularities and Enhanced Gauge Symmetries}
, hep-th/9605200}.).
At any rate we will find evidence for the above conjecture when
we compare the predictions based on the tensionless strings
and the geometry of the del Pezzo $\bb_d$.

\subsec{Geometry of the del Pezzo Surfaces}

In the following we describe some aspects of the geometry involving
the vanishing 2- and 4-cycles associated
to a del Pezzo Surface sitting in the Calabi-Yau
threefold. Although the properties of
the  tensionless strings associated to them are governed by the
local geometry, the choice of an appropriate global embedding
will be important in order to be able to
extract important physical quantities such as the number
of BPS states. In order to do that we have to identify the homology
of the vanishing 4-cycle within that of a given Calabi--Yau threefold.

The relevant del Pezzo surfaces\foot{See
\ref\demazure{M. Demazure, in {\sl S\'eminaire sur les Singularit\'es 
des Surfaces},
Lecture Notes in Mathematics {\bf 777}, Springer Verlag 1980, 23}
and 
\ref\hartshorne{R. Hartshorne, {\sl Algebraic Geometry},
Springer New York (1977)}, Chap. V, Sect. 4 for a
thorough exposition of this construction.} $\bb_k$ can be constructed by
blowing up $k$ generic points $P_i$ on $\IP^2$
as $i$ runs from 1 to $k$ where
 $1\le k\le 8$.
The divisor classes of $\bb_k$ are thus the class of lines $l$ in $\IP^2$
and the $k$ exceptional divisors $D_i$ lying above the points $P_i$, whose
class we denote by $e_i$.
The number of nontrivial homology elements is $h^{0,0}=h^{2,2}=1$ and
$h^{1,1}=1+k$, which gives $k+3$ as the  the Euler number.
The non zero intersections in the $(l,e_1,\ldots,e_k)$ basis
of $H^{1,1}(\bb_k)$ are $l^2=1$ and $e_i^2=-1$; moreover the anti-canonical
\def\kb{\bar{\cal K}}
class $\kb_k$ is $\kb_k=c_1(\bb_k)=3l-\sum_{i=1}^k e_i$.

A curve $\cC$ in the homology class 
$a l- \sum_{i=1}^k
b_i e_i$ intersects the line $l$ $a$ times and moreover passes
$b_i$ times through the points
$P_i$. Its degree is
\eqn\degree{d_{\cC}=\kb_k \cdot  \cC=3 a-\sum_{i=1}^n b_i}
and its arithmetic genus can be obtained from the Pl\"ucker
formula~\ref\gh{P. Griffith and J. Harris, {\sl Principles of Algebraic
Geometry}, J. Wiley \& Sons, New York (1978)} as
\eqn\genus{g_{\cC}={(a-1)(a-2)\over 2}-\delta=
1+{1\over 2}(\cC\cdot \cC-\kb_k\cdot \cC),}
where $\delta$ is the number of nodes (ordinary double points).
At each point $P_i$ one has $b_i(b_i-1)/2$ nodes and we
assume the curves to be otherwise smooth. If there are $\delta'$
($\kappa$) additional nodes (cusps) outside the $P_i$ \genus\ is reduced
by $\delta'$ ($\kappa$), hence \genus~is merely an upper
bound on the genus, known as arithmetic genus.

To  count the number of irreducible curves of genus $0$ note that for
such curves $b_i\le a$ as an irreducible curve of degree
$a$ in $\IP^2$
cannot pass more then $a$ times through any given point.
To obtain the number of lines one enumerates therefore
solutions of \degree,\genus \
with $d_{\cC}=1$, $g_{\cC}=0$ and $b_i\le a$. E.g. for $\bb_8$ one
finds seven classes  $(a;b_1,\ldots,b_8)=
(0;-1,0^7)$, $(1;1^2,0^6)$, $(2;1^5,0^3)$, $(3;2,1^6,0)$,
$(4;2^3,1^5)$, $(5;2^6,1^2)$ and $(6;3,2^7)$. Counted with the obvious
multiplicities due to the permutation of the points $P_i$
these are $240=8+28+56+56+56+28+8$ lines.
By the same method the number of lines on the other $\bb_k$
$i=7,\ldots,1$ turns out to be $56,27,16,10,6,3,1$.
A thorough counting of rational curves on $B_6$ has been performed
in ref.  \ref\itz{
C. Itzykson,
{\sl Counting rational curves on rational surfaces}, Int.J.Mod.Phys.
$\us{B8}$ (1994) 3703;\br
P. Di Francesco and C. Itzykson, {\sl Quantum intersection rings},
hep-th/9412175}  under the requirement that the curves
pass through a sufficient number of points to avoid the problem
of continuous moduli.

The fact that the lines are in
representations of the Weyl groups of $E_k$
for $k\ge 3$ \demazure\
can be recognized by looking at the classes ${\cal E}\in H^{1,1}(\bb_k)$
fulfilling $\kb_k\cdot {\cal E}=0$. These can be generated by
${\cal E}_1=l-e_1-e_2-e_3$ and  ${\cal E}_i=
e_i-e_{i+1}$, which span the root lattice of
$E_k$. Therefore one has an action of the Weyl group on $H^{1,1}(\bb_k)$
and curves $\cC$ of given degree $d_{\cC}$ must be organized
representations of the Weyl group.

The del Pezzo surface $\bb_9$ can be understood as the blow
up of $\IP^2$ at the nine intersection points of two cubic curves;
this defines an elliptic fibration over $\IP^1$ which can be described
generically by the equation
\eqn\bnine{
y^2 = x^3+x f_4(\zp)+ g_6(\zp)
}
where $\zp$ is the coordinate on the base $\IP^1$ and
$y,x$ the coordinates on the fibre. There  are 12
singular fibres above the discriminant locus $\Delta=4f_4^3+27g_6^2=0$;
therefore the Euler characteristic $\chi$ is  12.  Roughly speaking this is
the structure of half a K3; in that case $f_4,g_6$
are replaced by polynomials of the double degree $f_8,g_{12}$
leading to 24 singular fibres and two copies of $E_8$
in the intersection matrix.

There are $k+1$ real parameters associated to the volumes of the
independent holomorphic two cycles of $\bb_k$. However not all of
them will descend to K\"ahler moduli of the Calabi--Yau threefold $X$,
the actual number depends on the rank $r$ of the map $H^{1,1}(\bb_k)
\rightarrow H^{1,1}(X)$. To contract $\bb_k$ one has to shrink all the volumes
of its holomorphic 2-cycles to zero; in particular this means that
one has to restrict a codimension $r$ locus in the K\"ahler moduli space
of $X$. Moreover, if the class which measures the size of the
elliptic fibre of the del Pezzo is the same as that of the generic
fibre of the elliptically fibred three-fold, the contraction necessarily
collapses $X$ to its base $B$. Whereas this is the usual limit one
has to take in the F-theory compactification, this situation is clearly
inappropriate
in the context of M-theory compactifications to 5 dimensions.

In order to obtain contractions at codimension 1 in the
moduli space which do not change the dimension of $X$
as discussed before one has to perform
flops on 2-cycles in the del Pezzo which support K\"ahler classes of
$X$. This operation effectively contracts the exceptional divisors;
in particular a flop of an exceptional divisor $D_i$
results in a transition from $\bb_k$ to $\bb_{k-1}$. On the other hand if
the elliptic fibre of $\bb_k$ coincides with that of the elliptic fibration
$X$, as in the cases considered below, one has always to flop
the base $\IP^1$ because of the variation of the fibre type above the base.

\def\ch{\hat{C}}
An important aspect of the global embedding of the vanishing cycle
in the Calabi--Yau context is the fact that
 we can measure their volume,
and therefore the mass of the associated BPS state. Let $K=\sum J_i k_i$
denote the volume form of the Calabi--Yau threefold $X$, where $J_i$ are the
K\"ahler classes and $k_i$ special coordinates on the K\"ahler moduli
space and let $\ch_i$ denote the dual homology classes fulfilling
$J_i \ch_j = \delta_{ij}$. Classically the area of a cycle in the class
$\sum_i c_i \ch_i$ is then given by $\sum_i c_i k_i$.
In compactifications to four dimensions instanton
corrections may shift the position of the singularity associated to
a vanishing cycle away from the locus $\sum_i c_i k_i = 0$; as a
consequence multiples $n \ch_0$ of the class $\ch_0$ of
a vanishing cycle will have non-zero volume if this happens.
In particular, in order that the volumes of 2-cycles in a vanishing
4-cycles indeed all vanish, it will be important that there is no
such shift due to instanton corrections. The absence of
a shift in the five-dimensional case can be inferred from the independence
of the vector moduli space on the overall volume modulus of the
Calabi--Yau manifold, which sits in a hypermultiplet and scales the
action of the worldsheet instantons
\ref\bbs{K. Becker, M. Becker and A. Strominger,
{\sl Fivebranes, Membranes and Nonperturbative
String Theory}, \nup456(1995)130}.
In the four--dimensional
type IIA compactification one has to establish the coincidence of the
classical volume with the exact vanishing period to exclude a quantum
split of the areas associated to multiples of a given class.

In order to be a valid F-theory compactification we will choose our
Calabi--Yau threefold $X$ to be
elliptically fibred; to make contact with the heterotic picture we will
further require it to be a $K3$ fibration. As discussed previously
these restrictions are not necessary
conditions for the existence of the type of tensionless string transition
we consider, which requires only the local geometry being that
of a vanishing 4-cycle of the appropriate type and neither
a global elliptic or K3 fibred structure. Rather these restrictions
provide the appropriate global embedding for the local geometry
which is convenient for the counting of the BPS spectrum and the
physical interpretation. In particular we take the elliptic fibre of
the del Pezzo representing the vanishing 4-cycle to be that
of the elliptic fibred Calabi--Yau. This is only possible if the base of
the elliptic fibred Calabi--Yau is the Hirzebruch surface
$\ff_1$ \mvii. The Weierstrass model for the elliptic fibred three-fold
with base $\ff_n$ is defined by the equation
\eqn\ws{
E_8: y^2=x^3+x\sum_{k=-4}^{k=4}z^{4-k}f_{8-n k}(\zp)+
\sum_{k=-6}^{k=6}z^{6-k}f_{12-n k}(\zp)
}
where $n=1$ for $\ff_1$, $y,x$ are the coordinates on the elliptic fibre $E$,
$z$ is the coordinate on the fibre $F$ of $\ff_1$ and
$\zp$ the coordinate on its base $B$. Note that $F$ is the base of
a elliptic fibred K3, fibred itself over the base $B$ and that
the elliptic fibre is described by the simple elliptic singularity
$\IP^{1,2,3}[6]$ of $E_8$ type.

To extend the global description to the del Pezzo
surfaces $\bb_d$, $d=6,7$, consider the local form of the
singularity of the vanishing 4-cycle in $C^4$ \ref\reid{M. Reid in Journ\'ees
de G\'eometrique Alg\'ebraic d' Angers, Juillet 1979,
Sijthoff \& Noordhoff (1980) 273}:
\def\zpp{z^{\prime\prime}}
\eqn\lsing{\eqalign{
E_8 &: y^2=x^3+x^2f_2+xf_4+f_6\cr
E_7 &: y^2=x^4+x^3f_1+x^2f_2+xf_3+f_4\cr
E_6 &: \sum y^kx^lf_{3-k-l}=0 \ ,\cr
}}
where $f_n$ are homogeneous polynomials of degree $n$ in two variables
$\zp$ and $\zpp$.
The elliptic fibre defined by setting $\zp$ and $\zpp$
constant is no longer of the generic type for $d < 8$ but
corresponds to a symmetric torus. In other words, if we insist to
keep the elliptic fibre of the del Pezzo to be identical to that of the
Calabi--Yau fibration we have to consider elliptic fibred
threefolds with the corresponding fibres of $E_7$ and $E_6$
type, $\IP^{1,1,2}[4]$ and $\IP^{1,1,1}[3]$, 
respectively\foot{Calabi--Yau threefolds of this type have been 
discussed recently also in 
\ref\ibaI{G. Aldazabal, A. Font, L. E. Ibanez and A. M. Uranga,
{\sl New branches of string compactifications and their F-theory
duals, hep-th/960712}}.}.

After choosing
a section and restricting to a patch, the defining equation of the elliptic
fibred threefold with base $\ff_n$ replacing \ws\  becomes:
\def\wp{w^\prime}
\eqn\wsII{\eqalign{
E_7:\ y^2&=
x^4+x^2 \sum_k f_{4-nk}z^{2-k}w^{2+k} \cr
&+x\sum_k f_{6-nk}z^{3-k}w^{3+k} + \sum_k f_{8-nk}z^{4-k}w^{4+k}\cr
E_6:\ y^3+x^3&=y\sum_{k,l}x^lw^{2-l+k}z^{2-l-k}f_{4-2l-nk}
\cr &+x\sum w^{2+k}z^{2-k}f_{4-nk}+\sum w^{3+k}z^{3-k}f_{6-nk}
}}
where $f_n$ are homogeneous polynomials of the base variables $\zp,\ \zpp$.
The terms which appear in \wsII\ are restricted to respect the
$C^\star$ symmetries
$$\eqalign{
E_8:&(6-3n,4-2n,1,1,-n,0),\ (6,3,0,0,1,1)\cr
E_7:&(4-2n,2-n,1,1,-n,0),\ (4,2,0,0,1,1)\cr
E_6:&(2-n,2-n,1,1,-n,0),\ (2,2,0,0,1,1)\cr}
$$
acting on $(y,x,\zp,\wp,z,w)$, where $y,\ x$ are the coordinates
of the elliptic fibre, $\zp,\ \wp$ those of the base of $\ff_n$ and
$z,\ w$ those of the fibre of $\ff_n$. In this way one obtains
series of Calabi--Yau threefolds with the corresponding elliptic
fibres and base $\ff_n$. Most of them have a simple representation
in terms of hypersurfaces in weighted projective spaces
$$\eqalign{
E_8:&\IP^{1,1,n,4+2n,6+3n}[12+6n]\cr
E_7:&\IP^{1,1,n,2+n,4+2n}[8+4n]\cr
E_6:&\IP^{1,1,n,2+n,2+n}[6+3n]\ }
$$
respectively; in general the
appropriate description is in terms
of toric varieties as described in the appendix.
For $\ff_1$ one recognizes
the singularities \lsing\ as the local equations in the neighborhood of
$\zp=\zpp=0$.
The elliptic fibred Calabi--Yau threefolds obtained in this way have
hodge numbers $(h^{1,1},h^{2,1})=(4,148)$ and  $(5,101)$, respectively.
As described in \mvii\wittra\ these fibrations based on $\ff_1$ have
Higgs branches which describe F-theory on an elliptically fibred threefold
with base $\IP^2$ and hodge numbers related to the one with base $\ff_1$
by \mvii
$$
h^{1,1}=h^{1,1}-k,\qquad h^{2,1}= h^{2,1} + c_d-k
$$
In the present case we have $k=1$ and therefore the spectrum
on $\IP^2$ is expected to be $(3,165)$ and $(4,112)$, respectively;
it is straightforward to check that the transition
leads to corresponding elliptic
fibred threefolds are described by hypersurfaces in
$\IP^{1,1,1,3,6}$ and $\IP^{1,1,1,3,3}$.

In fact one can show using methods similar to \ref\bkkm{P. Berglund,
S. Katz, A. Klemm and P. Mayr,
{\sl New Higgs Transitions between Dual N=2 String Models}, hep-th/9605154}
that the three Calabi--Yau manifolds $\xthree_{(3,243)}$, $\xfour_{(4,148)}$
and $\xfive_{(5,101)}$ are connected by extremal transitions
where precisely the right number of rational curves are blown down
to provide the new hypermultiplets on the side with the larger number
of complex structure moduli.  From the viewpoint of F-theory, in the
six dimensional limit, where the size of the elliptic fiber
goes to zero, the $(4,148)$ model and $(5,101)$ model
can be viewed as a subset of the $(3,243)$ model where we have
tuned the complex structure on the threefold in a particular way.
To make a transition we have to give the elliptic fiber a finite size
which is only allowed if we further compactify on a circle
to 5 dimensions.  This is the geometrical
realization of turning on the Wilson
lines on the circle.
A more detailed description of these manifolds for the
$E_7$ and $E_6$ cases as toric varieties is given in the appendix.

\newsec{Counting holomorphic curves with mirror symmetry}

We are interested in BPS states, which arise from wrapping
membranes around supersymmetric two cycles. They correspond
to the holomorphic curves in the Calabi-Yau manifold $X$ \ref\bbs.
Luckily the techniques to compute the number
of such curves is available thanks to mirror symmetry!  The reader
should note that we are not discussing the physics of four dimensional
compactifications of type IIA on the manifold $X$, but rather we are using
the information gained by studying such conformal theories to count
the number of holomorphic 2-cycles in $X$; these holomorphic
2-cycles are relevant as BPS states even if we are talking about the
5-dimensional theory obtained by compactifying M-theory on $X$.
We will focus on the case of BPS states represented by genus zero
curves which are to be obtained
from the periods integrals on the mirror manifold of $X$ extending
the approach of \ref\cogp{P. Candelas, X. de la Ossa, P. Green and
L. Parkes, {\sl A Pair of Calabi-Yau manifolds as an exactly soluble
                  superconformal theory}, \nup359(1991)21}.  One can
naturally ask whether the BPS states
are represented by higher genus curves?
This for instance is naturally the case for holomorphic 2-cycles
in $K3$ which are represented by higher genus curves
\ref\bsv{M. Bershadsky, V. Sadov and C. Vafa,
{\sl D-Branes and Topological Field Theories}, \nup463 (1996) 166 }.
However even in that case it has been shown in
\ref\zy{S.-T. Yau and E. Zaslow,
{\sl BPS States, String Duality, and Nodal Curves on K3}, 
\nup 471 (1996) 503, hep-th/9512121}
that if one allows genus 0 curves with nodal singularities, it will in effect
have the full informations about the higher genus cases as well.
The results we shall find below by studying the genus 0 curves in the case
of del Pezzo also seem to take account of such singular maps automatically
and thus in effect contain the information about the higher genus
curves as well.

There is also the question of whether the holomorphic maps
come in isolated sets or are part of a continuous moduli space.
In the latter case
 mirror map computes
the Euler class of an appropriate bundle  (of `anti-ghost zero modes')
on the moduli space of holomorphic
cycles.  If the bundle coincides with the tangent bundle of moduli
space, this would simply give the Euler class of moduli space, which
can be interpreted as the `net number' of BPS states (
the notion of `net number' can be defined taking into account
that some can pair up to become non-BPS representations).  We
have not proven that when we have a family of holomorphic
curves, mirror map computes the relevant topological number
for the net number of BPS states, but in some cases
that we could check this, it turned out to be so,
which leads us to speculate about the general validity of such a statement.

First consider the Calabi-Yau $\xthree$ with hodge numbers
$(h^{1,1}=3,h^{2,1}=243)$,
which is an elliptic fibration with a single section over the
Hirzebruch surface $\ff_1$. This Calabi-Yau has two geometrical
phases, which we have discussed in the previous sections and
will also be described explicitly in terms of toric geometry
in the appendix .

In the first phase the manifold is a $K_3$ fibration and
the K\"ahler cone is spanned by  three classes: i) $C_F$ the
fibre $\IP^1$ of the $\ff_1$, ii) $C_D$ the exceptional section
$\IP^1$ of the $\ff_1$
and iii) $C_E$, a curve in the elliptic fibre. The Gromov-Witten
invariants of rational curves, i.e.
the number of holomorphic spheres with degree $d_F,d_D,d_E$ w.r.t.
these classes, $n_{d_F,d_D,d_E}$, can be counted (modulo the subtleties
noted above) using mirror symmetry.
As discussed before
the relevant classes for the counting of states of the tensionless string
are $C_D$ and $C_E$ and we will sometimes denote the relevant
$n_{0,d_D,d_E}$ by $n_{d_D,d_E}$.
 Here we have the following invariants.

{\vbox{\ninepoint{
$$
\vbox{\offinterlineskip\tabskip=0pt
\halign{\strut\vrule#
&\hfil~$#$
&\vrule#&~
\hfil ~$#$~
&\hfil ~$#$~
&\hfil $#$~
&\hfil $#$~
&\hfil $#$~
&\hfil $#$~
&\hfil $#$~
&\hfil $#$~
&\vrule#\cr
\noalign{\hrule}
& && d_E& 0 &  1  &  2  &  3 & 4 & 5 & 6&\cr
\noalign{\hrule}
&d_D &&  & &     &      &       &        &         &        &\cr
&0   &&  & &  480&   480&    480&     480&      480&     480&\cr
&1   &&  &1&  252&  5130&  54760&  419895&  2587788& 13630694&\cr
&2   &&  & &     & -9252& -673760& - 20534040&-389320128&-5398936120&\cr
&3   &&  & &     &      &848628&    115243155&6499779552& 219488049810&\cr
&4   &&  & &     &      &      &   -114265008& - 23064530112&-1972983690880&\cr
\noalign{\hrule}}
\hrule}$$
\vskip-7pt
\noindent
{\bf Table 1}: Invariants of $\xthree$ for rational curves with degree
$d_F=0$.}
\vskip7pt}}

Note that in terms of the tensionless string
description $d_D$ and $d_E$ denote the winding number
and the momentum quantum of the nearly tensionless string
wrapped around the circle.  The first line of this table is special in
that  $d_D=0$ corresponds
to no winding of tensionless strings.  So this part is just
the Kaluza-Klein momentum excitations on circle of the corresponding
massless states in the 6-dimensional theory.  Given
the fact that the net number of hypermultiplets minus
the vector multiplets (in 4-dimensional terms) is $-\chi/2$
and that for the manifold with $(h^{11},h^{21})=(3,243)$
we have $-\chi =480$ we get a perfect match with the corresponding
computation from the mirror map\foot{Note that the factor of 2
is there to make a full hypermultiplet on the supersymmetry side
(which for $d_D\not=0$ is effected by the negative reflection of $d_D$).}.
  This is a case where in fact
one can show that the moduli space of holomorphic curves
is not isolated and is in fact a copy of the Calabi-Yau itself
making us gain faith in the meaning of the numbers computed by
the mirror map.  Note that (because of the appearance of the Euler
number) this first row of the table is not
universal and depends on which Calabi--Yau we used to realize
this transition.  This is {\it not} the case
with the other rows in the table, corresponding to
non-vanishing winding
states of the tensionless strings
which turn out to be universal. In fact we have checked that
in the class of elliptic fibred threefolds whose elliptic fibre
is at the same time the fibre of the del Pezzo (as is necessary for
the above interpretation of the BPS states),
independently of how the corresponding transition is embedded in
the Calabi-Yau, these are unaffected \foot{The universal
invariants of the local geometry of the vanishing 4-cycles
without the above mentioned restriction are discussed in the
appendix.}.

Now we come to the more interesting numbers in the above table.
First of all note that for the singly wound state $d_D=1$
with
no momentum $d_E=0$ we have one BPS state but for all $d_D>1$
with $d_E=0$ there are none.  This implies
that the multiply wound tensionless string
with no momentum does not form a bound state.  For the winding
number one and momentum
quantum $k$ we can  read off the spectrum of BPS states from the
second line of the
above table and is precisely given by
\eqn\resI{\eqalign{
\hat \Lambda_{E_8}&={1 \over 2}\sum_{\alpha=even}
{\theta^8_\alpha(\tau)\over q^{-{1\over 2}}\eta(\tau)^{12}}
\cr &=
1+252 q +5130 q^2+54760 q^3+419895q^4 + 2587788 q^5 +
\ldots}}
Here $\theta_\alpha$ are the Jacobi Theta functions
and the $\eta$ is the Dedekind eta-function
$$\eqalign{
{1\over 2}\sum_{\alpha=even}
\theta^8_\alpha(\tau)&=1 + 240 q+ 2160 q^2 + 6270 q^2 + 17520 q^4+
30240 q^5+\ldots \cr
q^{1\over 2}\eta(\tau)^{-12}&=1+12 q + 90 q^2+520 q^3+2535 q^4+10908 q^5+\ldots
}$$
We have checked the agreement of these functions and the coefficients of
the instanton expansions up to 12th order in $q$.
This result is in perfect accord with expectations based
on tensionless strings discussed in section 2.

Note that we actually
have more information here for multiple
winding states.    Consider for example the third row
in the above table
which corresponds to the BPS states with double winding numbers
of the tensionless string.
  First of all, the fact that the numbers are
negative implies that we are dealing with moduli space of holomorphic
maps rather than isolated numbers.   We should thus interpret
these number, as in the first row, as the net number of BPS states.
 The net number of double wound states exhibit
the following quadratic relation in terms of the single wound states
$$
8\ n_{2,k}=-(k-1)\sum_{i=0}^k n_{1,k-i}\
n_{1,i}-\delta_{(k\ {\rm mod} \ 2),0}\
n_{1,{k\over2}}.
$$
The interpretation of this sum rule is a very interesting
one for which we do not know the answer (though we present
some speculations in the next section).  For example the first
non-vanishing number $-9252$ in the third row can be viewed as
$$
8(9252)=(252\cdot 252)+2(1\cdot 5130)+252$$

We also expect, though have not checked, that the higher
winding number BPS states can also be factorized in terms
of other BPS degeneracies with total winding number adding
up to the winding number of BPS state in question.

\subsec{New Physics at the Second Transition}
As discussed in section 3 the interesting transition,
corresponds to the second point where the class $k_D+k_E=0$.
In particular at this point the $252$ states corresponding
to $d_D=d_E=1$ become massless.  As discussed in section
2 the quantum number of these states include fields with
spin up to spin $3/2$.  We are thus seeing a massless
gravitino at this point, therefore suggesting enhanced local supersymmetry
at this transition point! This is novel in that this is happening
at a {\it finite} distance in moduli space (see also
\ref\other{M. Cvetic and D. Youm, {\sl BPS Saturated States
and Non-Extreme States in Abelian Kaluza-Klein Theory and
Effective N=4 Supersymmetric String Vacua}, hep-th/9508058\semi
I. Antoniadis, H. Partouche and T.R. Taylor, {\sl Spontaneous
Breaking of N=2 Global Supersymmetry}, \plt 372 (1996) 83, hep-th/9512006\semi
S. Ferrara, L. Girardello and M. Poratti, {\sl Spontaneous
Breaking of N=2 to N=1 in Rigid and Local Supersymmetric Theories},
\plt 376 (1996) 275, 
hep-th/9512180\semi
R. Rohm, {\sl Spontaneous supersymmetry breaking in supersymmetric
string theories}, \nup237 (1984) 553\semi
I. Antoniadis, C. Munoz and M. Quiros, {\sl Dynamical supersymmetry breaking
with a large internal dimension}, \nup397 (1993) 515\semi
E. Caceres, V. Kaplunovsky and M. Mandelberg, {\sl
Large volume limit in string compactifications, revisited}, hep-th/9606036\semi
E. Kiritsis, C. Kounnas,  M. Petropoulos and J. Rizos, {\sl
Solving the Decompactification Problem in String Theory}, 
\plt 385 (1996) 87, hep-th/9606087}).
 This is quite remarkable.
One may wonder whether we can prove that the corresponding BPS
state is stable.  Before we approach the first
transition point any BPS state
with a fixed $(d_D,d_E)$ has
 the same mass as the sum of masses of BPS states whose
$(d_D^i, d_E^i)$ add up to it.   So in principle there is a channel
were they could have decayed.  We do not believe this is the case
and believe that they form bound states at threshold;  for example
 the fact that the mirror map does not predict any BPS
state for $d_D>1,d_E=0$ already suggests that mirror map,
which counts only primitive instantons
(since we have subtracted multi-cover contributions) only counts
states which are stable.   At any rate we can actually make this rigorous
at least for the 252 states in the second transition.  In principle
the $252$ states can decay to the combination of $d_D=1,d_E=0$
states and $d_D=0,d_E=1$ states.
 However after the first transition
point where the sign of $k_D$ flips the situation changes.
In particular
the BPS state corresponding to $d_D=1,d_E=0$ at this second
transition has negative $mass$ but positive $(mass)^2$ and
so it is an ordinary massive BPS state \wittra .  Thus energetics
forbid the decay of the 252 states.  We are thus rigorously predicting
the existence of a massless stable particle with spin 3/2 at the
second transition point.  This is however not the end of the story.
In fact all the states with $(d_D,d_E)=(n,n)$  are massless
at the second transition point. Even though now we cannot prove
they are stable (as they can in principle decay to $n$ copies
of $(d_D,d_E)=(1,1)$ states) based on what we said above
we believe they represent BPS states which are bound states at threshold.  Thus
the massless
252 states are just the tip of the iceberg and we seem to see
an infinite tower of massless states, with net BPS degeneracies
$252,-9252,848628,...$,
 as had been anticipated
in \mvii \wittra .    As mentioned before we also
have a tensionless magnetic string at this point interacting
with all these massless states.  This suggests that the totality of
BPS state $252,-9252,...$ form some kind of ``representation''
for this new string.  The analogy we have in mind is that if
we have a gauge particle for some group $G$ it can interact
with massless particles which form non-trivial representations
of $G$.  Here we believe we have a non-critical string version
of this situation at hand where a non-critical tensionless
$E_8$ string interacts with the correspondingly
infinite tower of BPS states which can themselves be viewed
as negative tension $E_8$ strings wrapped around the circle
$n$ times with momentum $n$ around the circle.  This also suggests that
a quantum correction is responsible for shifting the tension
of the wrapped string from that of the unwrapped one.
Clearly there is a lot of new physics hidden here remaining
to be explored.  We cannot hesitate to speculate
about the implications of a better understanding of this
second transition point
for the question of supersymmetry breaking, given the fact
that we seem to have found as a `minor' part of the
infinitely many light BPS
states, a nearly massless spin 3/2 state!

\subsec{Other $E_d$}

Next we consider the elliptically fibred Calabi--Yau $\xfour$
with base $\ff_1$ and generic fibre of the type $P^{1,1,2}[4]$
with hodge numbers $(4,148)$. The four K\"ahler classes
are supported by the following classes of curves:
i) $C_F$, the fibre $\IP^1$ of $\ff_1$,
ii) $C_D$, the exceptional section of $\ff_1$,
iii) $C_E$, a curve in the elliptic fibre and iv) a new
class which will be denoted by $C_W$,
which is introduced by the second section
of the fibration. In fact this manifold is connected
to the Calabi--Yau threefold $\xthree$ by a transition
which contracts the $\IP^1$ of the new K\"ahler class,
$k_W=0$. From $n_{0,0,0,1}=96,\ n_{0,0,0,k}=0$ for
$k>0$ we see that there are 96 2-cycles contracted
by the transition, thus explaining the change of the number
of complex structure deformations $243-148+1=96$.
Moreover the Gromov--Witten invariants sum up as
$n^{(3)}_{i,j,k}=\sum_l n^{(4)}_{i,j,k,l}$ for $k_W=0$.

The rational curves $n^{(3)}_{0,1,k}$ generating the
partition function \resI\ split according to the $U(1)$
charges in the decomposition $E_8 \subset U(1) \times E_7$.
The relevant Gromov--Witten invariants $n^{(4)}_{n_F,n_D,n_E,n_W}$ are
shown in table 2.

{\vbox{\ninepoint{
$$
\vbox{\offinterlineskip\tabskip=0pt
\halign{\strut\vrule#
&\hfil~$#$
&\vrule#&~
\hfil ~$#$~
&\hfil ~$#$~
&\hfil $#$~
&\hfil $#$~
&\hfil $#$~
&\hfil $#$~
&\hfil $#$~
&\hfil $#$~
&\hfil $#$~
&\hfil $#$~
&\hfil $#$~
&\vrule#
&\hfil ~$#$~
&\vrule#\cr
\noalign{\hrule}
&  && d_W& 0 &  1  &  2  &  3 & 4 & 5 & 6& 7 & 8& 9 && \sum &\cr
\noalign{\hrule}
&d_E && &   &     &      &       &        &         &         &  &  & && &\cr
& 0 &&  & 1 &     &      &       &        &         &         &  &  & &&1&\cr
& 1 &&  & 1 & 56  &  138 &  56   &   1    &         &         &  &  &
&&252 &\cr
& 2 &&  &   &     &  138 &  1248 & 2358   &  1248   &  138    &  &  &  &&5130
&\cr
& 3 &&  &   &     &      &   56  & 2358   &  13464  &  23004  &  13464 & 2358 &
56 &&
54760 &\cr
& 4 &&  &   &     & &     &    1  &  1248 & 23004   &  103136 & 165117 &
103136  &&
419895 &\cr
\noalign{\hrule}}
\hrule}$$
\vskip-7pt
\noindent
{\bf Table 2}: Invariants of $\xfour$ for rational curves
with degree $d_F=0,\ d_D=1$.}
\vskip7pt}}

Note that the splitting of the $E_8$ states into $E_7$ states
is in perfect accord with the idea explained in section 2
in the context of tensionless strings and turning on Wilson
lines of $U(1)$ on the circle.  Moreover we should identify
the Wilson line of the $U(1)$ with the K\"ahler class
of $W$: $k_W=W$.  Also we have to identify $k_E+2 k_W={1\over R}$;
this is consistent with the fact that a curve of this type
is also in the elliptic fibre.

The states of the $\xthree$, which are multiply wound
around the circle, have a completely analogous group theoretical
decomposition into the classes of states of $\xfour$.
For example for the doubly wound states $d_D=2$ one finds

{\vbox{\ninepoint{
$$
\vbox{\offinterlineskip\tabskip=0pt
\halign{\strut\vrule#
&\hfil~$#$
&\vrule#&~
\hfil ~$#$~
&\hfil ~$#$~
&\hfil $#$~
&\hfil $#$~
&\hfil $#$~
&\hfil $#$~
&\hfil $#$~
&\vrule#
&\hfil ~$#$~
&\vrule#\cr
\noalign{\hrule}
& && d_W
        & 2    &  3  &  4   &  5    &     6   &  7       && \sum &\cr
\noalign{\hrule}
&d_E && &     &      &      &       &         &          && &\cr
& 2 &&  & -272& -2272&-4164 &-2272  &-272     &          &&-9252&\cr
& 3 &&  &     & -2272&-38088&-165600&-261840  &-165600   &&-673760&\cr
& 4 &&  &     &      &-4164 &-165600&-1484256 &-4961952  &&-20534040&\cr
\noalign{\hrule}}
\hrule}$$
\vskip-7pt
\noindent
{\bf Table 3}: Invariants of $\xfour$ for rational curves of degree $d_F=0,\
d_D=2$.}
\vskip7pt}}

Interestingly one can identify the following symmetries of the 
instanton numbers suggesting a kind of T-duality of the tensionless
string:
\eqn\dsym{\eqalign{
k_E \to k_E + 4 k_W,\ k_W \to -k_W:& 
\ \ {1 \over R} \to {1 \over R},\ W \to -W\cr
k_E \to -k_E,\ k_W \to k_W+k_E,\ k_D \to k_D+k_E:&
\ \ {1 \over R} \to {1 \over R},\ W \to -W+{1\over R},\cr
&\ RT \to RT + {1\over R} -2 W
}}
Whereas the first symmetry related to a Weyl symmetry of the
underlying $E_8$ is present for all values of $n_F$,
the second, more remarkable one, holds only for $n_F=0$. 
It points to a duality symmetry of the enlarged charge lattice
$\Gamma^{1,9}$ (note however that we have found no indication of a 
$R\to 1/R$ symmetry as is expected from the zero tension
limit). We expect the symmetries of the instantons to correspond to
monodromies of the periods around singular loci in the moduli space.

The $E_6$ case is realized in the Calabi--Yau threefold where the
generic elliptic fibre over the $\ff_1$
base is of type $P^{1,1,1}[3]$; its Hodge numbers are $(5,101)$.
The five K\"ahler classes supported by the classes $C_F,\ C_D,\ C_E$ as well
as two additional classes which we denote by $W_1$ and $W_2$.
With respect to these classes the invariant of the $E_8$ case split
according to the decomposition $E_8 \subset E_6\times U(1)_1 \times U(1)_2$
as shown in table 4.\br

{\vbox{\ninepoint{
$$
\vbox{\offinterlineskip\tabskip=0pt
\halign{\strut\vrule#
&\hfil~$#$
&\vrule#&~
\hfil ~$#$~
&\hfil ~$#$~
&\hfil $#$~
&\hfil $#$~
&\hfil $#$~
&\hfil $#$~
&\hfil $#$~
&\hfil $#$~
&\vrule#
&\hfil ~$#$~
&\vrule#\cr
\noalign{\hrule}
&  && d_{W_1}
        & 0    &  1  &  2   &  3    &     4 &  5    &   6     && \sum &\cr
\noalign{\hrule}
&d_{W_2}
    &&  &     &      &      &       &       &       &         && &\cr
& 0 &&  &    1&      &      &       &       &       &         &&1&\cr
& 1 &&  &    1&    27&   27 &      1&       &       &         &&56&\cr
& 2 &&  &     &      &   27 &     84&     27&       &         &&138&\cr
& 3 &&  &     &      &      &      1&     27& 27    &     1   &&56&\cr
& 4 &&  &     &      &      &       &       &       &     1   &&1 &\cr
\noalign{\hrule}}
\hrule}$$
\vskip-7pt
\noindent
{\bf Table 4}: Invariants of $\xfive$ for rational
curves of degree $d_F=0,\ d_E=1,\ d_D=1$.}
\vskip7pt}}

{\vbox{\ninepoint{
$$
\vbox{\offinterlineskip\tabskip=0pt
\halign{\strut\vrule#
&\hfil~$#$
&\vrule#&~
\hfil ~$#$~
&\hfil ~$#$~
&\hfil $#$~
&\hfil $#$~
&\hfil $#$~
&\hfil $#$~
&\hfil $#$~
&\hfil $#$~
&\vrule#
&\hfil ~$#$~
&\vrule#\cr
\noalign{\hrule}
&  && d_{W_1}
        & 2    &  3  &  4   &  5    &     6 &      7&      8&& \sum &\cr
\noalign{\hrule}
&d_{W_2}
    &&  &     &      &      &       &       &       &       && &\cr
& 2 &&  &   27&    84&    27&       &       &       &       &&138&\cr
& 3 &&  &     &    84&   540&    540&     84&       &       &&1248&\cr
& 4 &&  &     &      &    27&    540&   1224&   540 &     27&&2358&\cr
\noalign{\hrule}}
\hrule}$$
\vskip-7pt
\noindent
{\bf Table 5}: Invariants of $\xfive$ for rational
curves of degree $d_F=0,\ d_E=2,\ d_D=1$.}
\vskip7pt}}

The results are once again in perfect agreement
with the splitting of states based on Wilson lines
the identifcation being $k_E+2k_{W_2}+3k_{W_1}={1\over R}$,
$k_{W_1}=W_1$, $k_{W_2}=W_2$ and $k_D=RT$. There
are similar symmetries of the instanton numbers as in the $E_7$
case, moreover
there is again an analogous decomposition of the  multiple wound
states $d_D>1$ of $\xfour$ into those of $\xfive$.

\noindent{\bf Acknowledgement}:
We would like to thank D. Morrison, E. Verlinde,
E. Witten and S.-T. Yau for valuable discussions.
The research of C.V. is supported in part by NSF grant PHY-92-18167.
C.V. would also like to thank ICTP and CERN for hospitality while
this work was being completed. The research of A.K. was 
partially supported by the Clay Fund for Mathematics, 
through the Department of Mathematics, Harvard University.

\appendix{A}{Toric description of the threefolds $X^{(i)}$}

\def\br{\hfill\break}
\def\mp{\phantom{-}}

\def\tE{\theta_E}
\def\tF{\theta_F}
\def\tD{\theta_D}

In this appendix we will give the data which specify the  
Calabi-Yau threefolds as hypersurfaces in 
toric varieties. The toric varieties can be described by pairs of 
reflexive polyhedra $(\Delta,\Delta^*)$ in a four-dimensional lattice. 
The canonical hypersurfaces $(X,X^*)$ in the projective 
toric varieties $(\IP_\Delta,\IP_{\Delta^*})$
give rise to mirror pairs of Calabi-Yau threefolds
\ref\bat{V. Batyrev, 
{\sl Dual Polyhedra and Mirror Symmetry for 
Calabi-Yau Manifolds} alg-geom/9310004,
Journal Alg. Geom. 3 (1994) 493;
{\sl Variations of the mixed Hodge

Structure of affine hypersurfaces in toric
varieties}, Duke Math. Journal 69 (1993) 349}. The exact worldsheet instanton 
corrections on $X$ can be expressed in terms of the periods of the 
mirror manifold $X^*$.  Explicit formulas for the periods at points
of large radii are given in terms the topological data and the 
Mori generators of $X$ in\foot{We will adapt to the notation 
of \ref\hkty{S. Hosono, A. Klemm, S. Theisen and S.T. Yau,
\cmp167(1995)301,\nup433(1995)501}.} \hkty. 
The later data are most easily calculable from the dual 
polyhedron $\Delta^*$, which we give in the following.

For the $X^{(3)}$ case the the dual polyhedron is the convex hull
of the following points
\eqn\polyEVIII{\eqalign{
\nu^*_0&=[\mp 0, \mp 0, \mp 0, \mp 0],\quad
\nu^*_1=[ \mp 1, \mp 0, \mp 0, \mp 0],\quad 
\nu^*_2=[-1,-1,-6,-9], \cr 
\nu^*_3&=[\mp 0,\mp 1,\mp 0,\mp 0],\quad 
\nu^*_4=[\mp 0,    -1,-4,-6],\quad 
\nu^*_5=[\mp 0, \mp 0, \mp 1, \mp 0],\cr 
\nu^*_6&=[\mp 0, \mp  0,\mp  0, \mp 1],\quad
\nu^*_7=[\mp 0, \mp 0,-2,-3].}}
The hypersurface $X$ in the $\IP_{\Delta}$ can be represented
in the Batyrev-Cox variables \ref\batcox{V. Batyrev and D. Cox, {\sl
On the Hodge Structures of Projective Hypersurfaces in Toric
Varieties}, alg-geom/9306011}\ $x_0,\ldots,x_7$ 
as vanishing of the polynomial
\eqn\pol{P=x_0[x_5^3+x_6^2+x_7^6(x_4^{12}
(x_1^{18}+x_2^{18})+x_3^{12}(x_1^6+x_2^6))].}

The model exhibits two geometrical phases, which correspond to
two regular triangulations of $\Delta^*$ involving all points, and
one non-geometrical phase.  
The first geometrical phase admits a $K_3$-fibration 
as well as an elliptic fibration and the generators of the Mori cone are: 
$$\eqalign{\cr
           l^{(F)}&=[\mp  0; \mp  0, \mp 0,\mp  1, \mp 1,\mp 0,\mp 0, -2]\cr
           l^{(D)}&=[\mp 0, \mp 1,\mp  1,\mp  0, -1,\mp 0,\mp 0 ,-1]\cr
           l^{(E)}&=[-6; \mp 0, \mp 0, \mp 0, \mp 0, \mp 2, \mp 3, \mp 1],}$$
where the indices refer to the classes of the curves, which bound the dual 
vector in the K\"ahler cone (comp. sec. 4). 
The classical triple intersections and the 
integrals involving the second Chern class are 
$$\eqalign{ {\cal R} &= 8\, J_E^3 + 3\, J_E^2 J_F + 
J_E\, J_F^2 + 2\, J_E^2\, J_D + J_D\, J_E\, J_F\cr   
\int_M c_2 J_E& = 92, \qquad \int_M c_2 J_F = 36, \qquad 
\int_M c_2 J_D = 24}$$
The differential equations for the periods in the complex 
structure variables of $X^*$ are generalized hypergeometric 
systems\foot{Properties of these systems were studied intensively 
by Gel'fand-Kapranov-Zelevinskii \ref\gkz{I. M. Gel'fand, M. M. Kapranov, 
A. V. Zelevinskii, Funct. Anal. Appl 23, 2}.}. 
In the case at hand we have the differential operators 
($\theta:=z {d\over dz}$):
$$\eqalign{ 
{\cal L}_1 &= \tE(\tE - 2\, \tF - \tD) - 12(6\, \tE - 5)\,
(6\, \tE - 1 )\, z_E \cr
{\cal L}_2 &=  \tF\, (\tF - \tD) - ( 2\, \tF + \tD - \tE -2)\,
    (  2\, \tF + \tD- \tE-1)\, z_F\cr
{\cal L}_3 &= \tD^2 - (\tD -1 - \tF)\, (2\, \tF + \tD - \tE - 1)\, z_D}
$$
From the periods and the discriminant of $X^*$  
$$\eqalign{
\Delta_1=&(1-z_E)^3(1-z_E-z_E z_D)- z_E^2 z_F (8(1-z_E)^2-
16 z_E^2 z_F+36 z_E z_D \cr&- 36 z_E^2 z_D+27 z_E^2 z_D^2)  
\cr
\Delta_2=&(1 - 4 z_F)^2 - z_D + 36 z_F z_D - 27 z_F z_D^2.}
$$
one can calculate $F^g_{top}$ which enjoys an expansion in terms 
of the Gromov-Witten invariants for elliptic and higher genus curves
respectively \ref\bcov{M. Bershadsky, S. Cecotti, H. Ooguri and
C. Vafa, \nup405(1993)279,\cmp165(1994)311}. The behavior of 
$F^1_{top}$ near the dicriminants is $z_E^{-{26\over 3}}$,
$z_F^{-{4}}$, $z_D^{-3}$, $\Delta_1^{-{1\over 6}}$ and $
\Delta_2^{-{1\over 6}}$.  For the elliptic curves one obtains

{\vbox{\ninepoint{
$$
\vbox{\offinterlineskip\tabskip=0pt
\halign{\strut\vrule#
&\hfil~$#$
&\vrule#&~
\hfil ~$#$~
&\hfil ~$#$~
&\hfil $#$~
&\hfil $#$~
&\hfil $#$~
&\hfil $#$~
&\hfil $#$~
&\hfil $#$~
&\vrule#\cr
\noalign{\hrule}
&d_F=0 && d_E& 0 &  1  &  2  &  3 & 4 & 5 & 6&\cr
\noalign{\hrule}
&d_D &&  & &     &      &       &         &         &        &\cr
&0   &&  & &    4&      &       &         &         &        &\cr
&1   &&  & &   -2&  -510& -11780&  -142330& -1212930 & -8207984&\cr
&2   &&  & &     &  762 & 205320& 11361870&  31746948& 5863932540&\cr
&3   &&  & &     &      &-246788&-76854240&-6912918432&-32516238180&\cr
&4   &&  & &     &      &      &  76413073& 278663327760&348600115600&\cr
\noalign{\hrule}}
\hrule}$$
{\bf Table 6}: Gromov-Witten invariants for the genus one curves.
\vskip7pt}}

The second phase is connected to the first phase by a flop of the
class $D$, which is the $\IP^1$ in the del Pezzo, whose complexified 
size parameter in the $K_3$ phase was the modulus of the dilaton 
of the heterotic theory, i.e. $\tilde l^{(1)}=l^{(E)}+l^{(D)},\ 
\tilde l^{(2)}=l^{(F)}+l^{(D)},\ \tilde l^{(3)}=-l^{(D)}$

The topological data in this phase are given by
$$\eqalign{{\cal R} = 8\tilde J_1^3+3\tilde J_2&\tilde J_1^2+\tilde J_2^2
\tilde J_1+ 9 \tilde J_1^2\tilde J_3+3\tilde J_2\tilde J_3 \tilde J_1+
\tilde J_2^2\tilde J_3+9\tilde J_3^2\tilde J_1+
3\tilde J_2  \tilde J_3^2+9\tilde J_3^3\cr  
\int_M c_2 \tilde J_1& = 92, \qquad \int_M c_2 \tilde J_2 = 36, \qquad 
\int_M c_2 \tilde J_3 = 102}$$
The transition shrinking the 4-cycle corresponds to the limit 
$\tilde z_1=\infty$, $\tilde z_3=0$  where $\tilde z_1 \tilde z_3$ 
held fixed. This can be seen from the relation of the $\tilde l^{(i)}$
$$\bar l^{(1)}=\tilde l^{(1)}+\tilde l^{(3)}=
(-6; 0, 0,  1, 0, 2, 3 , 0),\quad       
  \bar l^{(2)}=\tilde l^{(2)}=
(\mp 0 ; 1,  1,  1, 0, 0, 0,-3)$$ 
to the Mori generators $\bar l^{(i)}$ of $X^{1,1,1,6,9}[18]$ model, whose
polyhedron $\Delta^*$ is given by the convex hull of the points
in \polyEVIII, with $\nu^*_4$ omitted. 
The analytic 
continuation from the point $\tilde z_i=0$ to the point 
$\hat z_i=0$ in the new variables 
$$\hat z_1={1\over \tilde z_1}, \quad
           \hat z_2=\tilde z_2, \quad
           \hat z_3=\tilde z_1\tilde z_3$$
is simplified by the fact that the expansions for five of the 
of the eight periods around the large complex structure limit 
converge also at the transition point. 
Furthermore the period related to the K\"ahler class $\tilde{k}_1$ 
and its dual are analytically continued into linear combinations
of the  
solutions $\omega^{1/6,0,0}\propto \hat z_1^{1 \over 6}+\ldots $ and $
\omega^{5/6,0,0} \propto \hat z_1^{5 \over 6 }+\ldots $, 
which vanish at $\hat z_i=0$; this proofs that the volumes
of the 2-cycle and 4-cycle vanish also after including the quantum 
corrections in the four-dimensional theory obtained by a type IIA
compactification.

The structure for the threefolds $\xfour$ and $\xfive$ is similar
and we omit the details.
The polyhedra $\Delta^*$ for $X^{(4)}$ and $X^{(5)}$ 
can be obtained by adding the point $\nu^*_8=(0,0,1,1)$ and
$\nu_9^*=(0,0,1,2)$ to the points in \polyEVIII\ respectively.   
For the $X^{(4)}$ we choose a phase in which the Mori generators 
${l'}^{(D)}, {l'}^{(E)},{l'}^{(F)}$ and ${l'}^{(W)}$
are related  to the  Mori generators of $X^{(3)}$ by 
$l^{(E)}={l'}^{(E)}+3 {l'}^{(W)}$, 
$l^{(D)}={l'}^{(D)}$ and 
$l^{(F)}={l'}^{(F)}$. Likewise the $X^{(5)}$ model exhibits 
a phase in which the transition to $X^{(4)}$ is apparent as the 
Mori generators  $\hat l^{(D)}$,  $\hat l^{(E)}$, $\hat l^{(F)}$, 
$\hat l^{(W_1)}$ and $\hat l^{(W_2)}$ are related to the one 
of the $X^{(4)}$ model by 
${l'}^{(E)}=\hat l^{(E)}+\hat l^{(W_1)} $,
${l'}^{(W)}=2\hat l^{(W_2)}+\hat l^{(W_1)}$,
${l'}^{(D)}=\hat l^{(D)}$ and
${l'}^{(F)}=\hat l^{(F)}$.

\appendix{B}{Universal structure of the 4-cycle singularity}
In the following we describe the structure
of the differential equations and their solutions which
govern the Gromov--Witten invariants of the local vanishing
4-cycle $\bb_k$ independently of the global embedding in the
Calabi--Yau. Specifically we study the dependence of
the mirror maps on the K\"ahler modulus of the collapsing
4-cycle. The restricted one modulus system extracts
the universal piece of the invariants associated to the
vanishing 2-cycle inside the 4-cycle.

The universal behavior of the singularity is completely
governed by a Mori vector $l $ which we can associate
to the local form \lsing. For $E_k,\ k=8,7,6,5$ one has
\eqn\moriI{\eqalign{
E_8:&\ l  = (-6|-1,3,2,1,1,0,\dots),\cr
E_7:&\ l  = (-4|-1,2,1,1,1,0,\dots),\cr
E_6:&\ l  = (-3|-1,1,1,1,1,0,\dots),\cr
E_5:&\ l  = (-2,-2|-1,1,1,1,1,1,0,\dots)
}}
This can be understood as follows: the rational curve $\hat{C}$ dual to $l $
is contained in the divisor $D_1:\{x_1=0\}$. The polynomial restricted to
$D_1$ has a $C^\star$ symmetry given by the remaining entries of $l $
which implies a form consistent with the local description of the
singularity \lsing.

Let $ k $ denote the relevant K\"ahler modulus and $ z = z ( k )$
the mirror map. From the vectors $l $ we obtain a differential operators
$\cL$ which govern
the mirror map $ z $
in the limit where all other volumes become large compared
to it. The differential operator $\cL$ turns out to be
closely related to the differential operator of the elliptic fibre
$\cL_{ell}$:
$$
\cL = \cL_{ell}\  \theta
$$
where $ \theta =  z {d \over d z }$ and $\cL_{ell}$ are
the differential operators of the elliptic curves
$P^{1,2,3}[6]$, $P^{1,1,2}[4]$, $P^{1,1,1}[3]$, $P^{1,1,1,1}[2,2]$, (see
e.g \ref\ly{B. Lian
and S.-T. Yau, Arithmetic properties of mirror Map and quantum
coupling, \cmp176 (1996) 163}):
$$\eqalign{
E_8:&\ \cL_{ell}=\theta^2-12z(6\theta+5)(6\theta+1)\cr
E_7:&\ \cL_{ell}=\theta^2-4z(4\theta+3)(4\theta+1)\cr
E_6:&\ \cL_{ell}=\theta^2-3z(3\theta+2)(3\theta+1)\cr
E_5:&\ \cL_{ell}=\theta^2-4z(2\theta+1)^2
}$$
Given a solution $\omega$ of $\cL$ we have also
a solution $\omega_{ell}= \theta\ \omega$ of $\cL_{ell}$. On the other
hand the fundamental period of $\cL$ \hkty:
$$
w_0( z ,\rho) = \sum_n c(n+\rho) z^{n+\rho}
$$
is given by the constant
solution, whereas the mirror map $ z $ is given by the single
logarithmic solutions. The expression for the instanton
corrected Yukawa couplings can be expressed in terms of
$w_0$ as 
$$
K( k )=\partial_k\partial_k\big({1\over 2 w_0}K^0\partial_\rho\partial_\rho
w_0|_{\rho=0}\big)( k )
$$
and subtracting multiple covers we obtain the
following Gromov--Witten invariants:\br

{\vbox{\ninepoint{
$$
\vbox{\offinterlineskip\tabskip=0pt
\halign{\strut\vrule#
&\hfil~$#$
&\vrule#&~
\hfil ~$#$~
&\hfil ~$#$~
&\hfil $#$~
&\hfil $#$~
&\hfil $#$~
&\hfil $#$~
&\hfil $#$~
&\vrule#\cr
\noalign{\hrule}
&    &&  d & 1 &  2  &  3  &  4 & 5 & 6 &\cr \noalign{\hrule}
&k   &&    &   &     &      &       &        &                 &\cr
&8   &&    &252&-9252&848628&-114265008&18958064400
&-3589587111852&\cr
&7   &&    & 56& -272&  3240& -58432 &1303840 &  -33255216       &\cr
&6   &&    & 27&  -54&   243& -1728  & 15255     &  -153576        &\cr
&5   &&    & 16&  -20& 48   &-192  & 960        & -5436    &\cr
&0   &&    &  3&   -6&    27&  -192  &      1695 &  -17064         &\cr
\noalign{\hrule}}
\hrule}$$
\vskip-7pt
\noindent
{\bf Table 6}: Invariants of the vanishing 4-cycles for $k=  0,5,6,7,8$.
\vskip7pt}}
\noindent
As suggested previously, these would correspond to the net degeneracies
of massless BPS states at the point where a 4-cycle
of type $E_d$ shrinks to zero size. 
Amusingly the $E_6$ series
turns out to be precisely 9 times the series one obtains for
$\IP^2$, shown in the last row, which would arise
for compactification of M-theory (or type IIA) on the
Z-orbifold \mvii \wittra .

\listrefs
\bye